\documentstyle[12pt]{article}
\def\be{\begin{equation}}
\def\ee{\end{equation}}
\def\ba{\begin{array}}
\def\ea{\end{array}}
\def\beqn{\begin{eqnarray}}
\def\eeqn{\end{eqnarray}}
\def\nonum{\nonumber}
\def\bt{\begin{tabular}}
\def\et{\end{tabular}}
\def\bc{\begin{center}}
\def\ec{\end{center}}

\begin{document}
 \title{Implications of texture 4 zero lepton mass matrices
for $U_{e3}$}
 \author{Monika Randhawa, Gulsheen Ahuja, and Manmohan Gupta \\
 {\it Department of Physics,} \\
 {\it Centre of Advanced Study in Physics,} \\
 {\it Panjab University,Chandigarh-160 014, India.}}   
 \maketitle
 \begin{abstract}
 Lepton mass matrices similar to  texture 4 zero quark
mass matrices, known to be quite
 successful in explaining the CKM phenomenology,
 have been considered  for finding the mixing matrix 
 element $U_{e3}\,(\equiv s_{13})$ respecting the CHOOZ constraint, 
 with $s_{12}$ and
 $\Delta m_{12}^2$ constrained by SNP and $s_{23}$ and $\Delta m_{23}^2$
 constrained by ANP. Taking charged lepton mass matrix $M_l$ 
to be diagonal,
 we find that the ranges of
 $s_{13}$ corresponding to different SNP solutions very well include
 the corresponding values of $s_{13}$ found by Akhmedov {\it et al.}
 by considering neutrino mass matrix $M_{\nu}$ with no texture
zeros. Considering $M_l$ and $M_{\nu}$  both  to be real and
non-diagonal, $s_{13}$ ranges for the four SNP
 solutions come out to be:
$ \sim 0 - 0.19~({\rm LMA}),~ 
 0.038 - 0.093~ ({\rm SMA}),~
  0.042 - 0.095~({\rm LOW}),~
0.038 - 0.096~({\rm VO}),$
which remain of the same order when $M_l$ and $M_{\nu}$
 are considered to be complex and non-diagonal. 
 \end{abstract}
The observation of neutrino oscillations by 
Super-Kamiokande (SK) \cite{sk}  
 as well as by Sudbury Neutrino Observatory (SNO) \cite{sno}
has provided  unambiguous signal for physics beyond the Standard
Model (SM).   
This essentially implies that the neutrinos are massive
non-degenerate particles and the 
 observed flavor eigenstates are linear combinations of mass 
 eigenstates, in parallel to the quark mixing phenomenon.
In this context, an intense amount of activity,
 both at the experimental as well as at the 
 phenomenological level, is being carried out to fix the parameters 
 of the neutrino oscillations 
   as well as the underlying textures of 
  the neutrino mass matrices  \cite{bil}-\cite{gloneut}.

    Several detailed and exhaustive analyses  have been 
 carried out in the case of the solar neutrino data as well as in the 
 case of the atmospheric neutrino data \cite{spec}. 
 These analyses along with the 
 results from several other experiments have provided valuable 
 information about the masses and the mixing parameters. 
 The constraints on masses and mixings are presented in terms of the 
 mixing angles $( \theta_{12}, \theta_{23}, \theta_{13})$, defined 
 analogous to the quark mixing angles, and the mass square 
 differences defined as 
 $\Delta m_{sol}^2 \equiv \Delta m_{12}^{2}= m_2^{2}-m_1^{2}$, 
 $\Delta m_{atm}^2 \equiv \Delta m_{23}^{2}= m_3^{2} 
-m_2^{2}$. 
The relationship between the mass eigenstates
 $(\nu_1,\,\nu_2,\,\nu_3)$ and the flavor eigenstates 
 $(\nu_e,\,\nu_{ \mu},\,\nu_{\tau})$ is expressed through the mixing
 matrix, for example
\be \left( \ba{c} \nu_e \\ \nu_{\mu} \\
\nu_{\tau} \ea \right)  = \left( \ba{ccc} U_{e1} &  U_{e2} &  U_{e3} \\ 
                     U_{\mu 1} &  U_{\mu 2} &  U_{\mu 3} \\ 
                     U_{\tau 1} &  U_{\tau 2} &  U_{\tau 3} \ea 
                     \right) \left( \ba{c} \nu_1 \\ \nu_2 \\
\nu_3 \ea \right). \label{ujk}  \ee
In the PDG representation \cite{pdg},
 the mixing matrix $U$  is given as
\be
U= \left( \ba{ccl} c_{12} c_{13} & s_{12} c_{13} & 
  s_{13}e^{-i \delta} \\ - s_{12} c_{23} - c_{12} s_{23} 
  s_{13} e^{i \delta} & c_{12} c_{23} - s_{12} s_{23} 
  s_{13} e^{i \delta} & s_{23} c_{13} \\ s_{12} s_{23} - c_{12} 
  c_{23} s_{13} e^{i \delta} & - c_{12} s_{23} - s_{12} c_{23} 
  s_{13} e^{i \delta} & c_{23} c_{13} \ea \right), \ee
 where $c_{ij}={\rm cos}\theta_{ij}$ and
   $s_{ij}={\rm sin}\theta_{ij}$ for
 $i,j=1,2,3$ and $\delta$ represents the CP violating phase. 

The best fit oscillation parameters 
for the  Atmospheric Neutrino Problem (ANP) are \cite{lisi} :
\be  \Delta m_{23}^{2} = 3.0 \cdot 10^{-3} {\rm eV}^2,  \qquad 
       {\rm sin}\,\theta_{23} = 0.71\,.     \label{ans}  \ee
  Similarly the global analysis
incorporating the SNO data  \cite{gloneut}, implies the
following ``best fit'' parameters
for  the various solutions of  Solar Neutrino
 Problem (SNP) : 
 \beqn 
 \Delta m_{12}^{2} = 8.0  \cdot 10^{-10}~\rm{eV}^{2}, & \qquad 
    {\rm sin}\,\theta_{12} =0.61\,, & \qquad {\rm VO}, 
  \label{vos} \\ 
\Delta m_{12}^{2} = 4.2 \cdot 10^{-5}\,~ \rm{eV}^{2}, & \qquad 
  {\rm sin}\,\theta_{12} =0.54\,, & \qquad {\rm LMA}, 
 \label{lmas} 
 \\ 
\Delta m_{12}^{2} =  5.0 \cdot 10^{-6}~\rm{eV}^{2}, & \qquad 
 {\rm sin}\,\theta_{12} = 0.025\,, & \qquad {\rm SMA}, 
  \label{smas} 
  \\ 
  \Delta m_{12}^{2} = 9.0 \cdot 10^{-8}~\rm{eV}^{2}, & \qquad 
     {\rm sin}\,\theta_{12} = 0.61\,, & \qquad {\rm LOW}. 
  \label{lows} 
  \eeqn 
 
 Although the above mentioned constraints are for two flavor oscillations, 
 however, in view of the CHOOZ constraint \cite{chooz} 
 \be  {\rm sin}^{2}\,\theta_{13} \equiv |U_{e3}|^{2} \leq (0.06-0.018)~ 
~~ {\rm for}~~~ \Delta m_{31}^{2}=(1.5-5) \times 10^{-3}~ {\rm eV}^2, 
  \label{cho}  \ee 
 these remain largely valid in the case of three flavor oscillations also.

    Among the several issues which need to be thoroughly examined 
 to enhance our understanding regarding the phenomenology of neutrino 
 oscillations is the issue of finding the lower limit of 
  $|U_{e3}|$, which in PDG representation corresponds to  
 $s_{13}$. 
 Because of the sensitive dependence on $s_{13}$ of the probabilities 
  of the long baseline (LBL) experiments and  sub-dominant 
 $\nu_e \rightarrow \nu_{\mu(\tau)}$ oscillations of atmospheric neutrinos, 
a knowledge of its value would be very helpful 
 in furthering our understanding of the neutrino oscillation 
 phenomenology. Besides this, a 
 knowledge of $s_{13}$ will be very important in distinguishing 
 various models of mass matrices and mixing schemes, as recently 
 emphasized by Barr and Dorsner \cite{bado}.

    The knowledge of mixing angles naturally leads to the question of 
    finding the underlying  mass matrices. 
This motivates one to go into specific schemes of these for leptons as 
the mixing matrix  can be related to the mass matrices. In this 
context,  texture specific mass matrices have been considered in 
the literature \cite{bado}-\cite{so10}
  which are able to 
explain several general features of the neutrino oscillation data 
 with good deal of success. However, 
the investigations do not go into the 
details of the implications of the texture structure 
 on $U_{e3}$.
 In view of the fact that
 see-saw mechanism 
  respects the texture structure \cite{frxing}  
as well as that texture specific matrices can be obtained in 
 the context of SO(10) grand unified theories \cite{so10}, 
  it is  desirable to carry out a detailed analysis of such structures.

A particular type of texture
4 zero quark mass matrices
 of the form 
\be 
 M_{U}=\left( \ba{ccc}0 & A_{U} & 0 \\ 
  A_{U}^{*} & D_{U} & B_{U} \\ 
  0 &  B_{U}^{*}  &  C_{U}  \ea \right), \qquad 
 M_{D}=\left( \ba{ccc}0 & A_{D} & 0 \\ 
  A_{D}^{*} & D_{D} & B_{D} \\ 
  0 &  B_{D}^{*}  &  C_{D}  \ea \right),
 \label{qmm} 
 \ee
 have shown a good deal of success in accommodating
 the Cabibbo-Kobayashi-Maskawa (CKM) phenomenology \cite{p1,frxing,p4}.
    Further, it has also been shown that such mass matrices could be
 generated from grand unified theories (GUTs) \cite{guts1} as well as
 that these are ``natural" in the sense of Peccei and Wang \cite{pewa}.
 Apriori
 there is no reason why the lepton mass texture and quark mass texture
 should be of the similar kind, nevertheless, it is interesting to see the
 consequences of a similar structure which may serve as guiding stone for
 the theories of neutrino mass matrices.

 The purpose of the present paper is to investigate
the implications of the   
texture 4 zero lepton mass matrices,  similar to the ones defined
 in equation (\ref{qmm}), for the 
 neutrino oscillation parameters.
 In particular, we intend to
 examine the implications of  
 these  matrices for $s_{13}$, keeping in mind the 
 constraints posed by the atmospheric neutrino data, various 
 solutions of the SNP as well as the results from CHOOZ. 
As a special case we have also investigated the implications
of the mass matrices wherein charged lepton mass matrix
is taken as diagonal, primarily for the sake of comparison of our results
with those of Akhmedov {\it et al.} \cite{akho}
as well as because of the recent
interest in such a case \cite{kang}-\cite{glashow}.
Further, it would be interesting to estimate the CP violating  
Jarlskog's rephasing invariant parameter \cite{jarl},
 even in the absence of observation of CP
violation in the leptonic sector.

  Therefore, to begin with, we consider lepton mass matrices which are 
similar to those given by equation  (\ref{qmm}).  Unlike 
 quark mass matrices where the $V_{\rm CKM}$ 
  matrix requires the elements 
 of any of the mass matrices to satisfy the hierarchy, 
 $|A| \ll |B| \simeq D < C$  \cite{p1}, 
 no such restriction is imposed on the lepton mass 
 matrices. Therefore, for the leptonic sector, mass matrices 
 $M_l$ and 
$M_{\nu}$ for the charged leptons and neutrinos  
respectively  are :
 \be 
 M_{l}=\left( \ba{ccc}  
0 &A _{l} & 0      \\ 
A_{l}^{*} & D_{l} &  B_{l}     \\ 
 0 &     B_{l}^{*}  &  C_{l} \ea \right), \qquad 
M_{\nu}=\left( \ba{ccc}  
0 &A _{\nu} & 0      \\ 
A_{\nu}^{*} & D_{\nu} &  B_{\nu}     \\ 
 0 &     B_{\nu}^{*}  &  C_{\nu} \ea \right), 
 \label{nmm6} 
 \ee 
where  $A_{l(\nu)} = |A_{l(\nu)}|e^{i\alpha_{l(\nu)}}$ 
and  $B_{l(\nu)} = |B_{l(\nu)}|e^{i\beta_{l(\nu)}}$. 
 As $M_l$ and $M_{\nu}$ are hermitian, these can be exactly diagonalized 
by using the  
unitary matrices $V_l$ and $V_{\nu}$, for example 
 \beqn 
 V_{l}^{\dagger}M_{l}V_{l} & = & {\rm diag}(m_{e},-m_{\mu}, 
  m_{\tau})\,,\\ 
 V_{\nu}^{\dagger}M_{\nu}V_{\nu} & = & {\rm diag}(m_1,-m_2, 
  m_3)\,.  \eeqn 
The corresponding lepton mixing matrix is given as 
 \be 
 U = V_l^{\dagger}V_{\nu} \,.
 \label{vlep} 
  \ee 

For    details of the diagonalizing transformations
we refer the reader to reference \cite{trans}.
Using the hierarchy of charged lepton masses, the diagonalizing
 transformation $V_l$ for $M_l$ can be simplified as
\be 
 V_l \cong \left( 
 \renewcommand{\arraystretch}{1.7} \ba{ccc} 
e^{i\alpha_l} &    
  -{\sqrt\frac{m_e}{m_{\mu}}}\,e^{i\alpha_l} &  
 {\sqrt \frac{m_e m_{\mu} (m_{\mu}+D_l)}{m_{\tau}^2(m_{\tau}-D_l)}}
\,e^{i\alpha_l}\\ 
   {\sqrt \frac{m_e(m_{\tau}-D_l)}{m_{\mu} m_{\tau}}}   
 &  {\sqrt \frac{m_{\tau}-D_l}{m_{\tau}}}   & 
 {\sqrt \frac{m_{\mu} + D_l}{m_{\tau}}}\\ 
 -{\sqrt \frac{m_e(m_{\mu} + D_l)}{m_{\mu} m_{\tau}}}\, e^{-i\beta_l} & 
 -{\sqrt \frac{m_{\mu}+D_l}{m_{\tau}}}\,e^{-i\beta_l} & 
 {\sqrt \frac{m_{\tau}-D_l}{m_{\tau}}}\,e^{-i\beta_l}  
 \ea \right). \label{mlapp} \ee 
 
In the absence of knowledge about absolute neutrino masses,
$V_{\nu}$ can not be simplified in the same manner. However,
some simplification can be achieved by noting that 
texture 4 zero  
mass matrices considered here
are not able to reproduce inverted mass hierarchy,
 primarily because of (1,1) elements in $M_l$ and
$M_{\nu}$ being zero.
Therefore, considering natural hierarchy,
we can consider
  $m_3 \gg m_2,m_1$. Hence the diagonalizing matrix $V_{\nu}$
for $M_{\nu}$ can be  simplified  as 
\be V_{\nu} \cong \left( 
 \renewcommand{\arraystretch}{1.7} \ba{ccc} 
 {\sqrt\frac{m_2}{m_1+m_2 }}\,e^{i\alpha_{\nu}} &     
  -{\sqrt\frac{m_1}{m_1+m_2 }}\,e^{i\alpha_{\nu}} &  
 {\sqrt \frac{m_1 m_2(m_2-m_1 +D_{\nu})} 
{m_3^2 (m_3-D_{\nu})}}\,e^{i\alpha_{\nu}}\\ 
   {\sqrt \frac{m_1(m_3-D_{\nu})}{m_3 
  (m_1+m_2)}}   
 &  {\sqrt \frac{m_2(m_3-D_{\nu})} 
{m_3(m_1+m_2)}}   & 
 {\sqrt \frac{m_2-m_1  + D_{\nu}}{m_3}}\\ 
 -{\sqrt \frac{m_1(m_2-m_1  + D_{\nu})} 
{  m_3(m_2+m_1)}}\,e^{-i\beta_{\nu}} & 
 -{\sqrt \frac{m_2(m_2-m_1  + D_{\nu})} 
{m_3(m_2+m_1)}}\,e^{-i\beta_{\nu}} & 
 {\sqrt \frac{m_3-D_{\nu}}{m_3}}\,e^{-i\beta_{\nu}}  
 \ea \right). \label{mnapp} \ee 
 
Using equation (\ref{vlep}) and the exact diagonalizing
transformations given in reference \cite{trans}, one can 
easily construct the corresponding lepton mixing matrix in terms of  
 the charged lepton masses, neutrino masses, $D_l$, $D_{\nu}$ and 
phases $\phi_1\,(=\alpha_l - \alpha_{\nu})$
 and $\phi_2\,(=\beta_l - \beta_{\nu})$. 
For the purpose of calculations we have used
 the exact expressions, however to facilitate
 the understanding of the dependence of the elements of the
 mixing matrix on various parameters, the simplified 
expressions for  $U_{e3}$, $U_{e2}$
and $U_{\mu 3}$, the 
elements of the mixing matrix directly related to the mixing angles
$\theta_{13}$, $\theta_{12}$ and $\theta_{23}$, are as follows:
{\small
 \beqn
U_{e3}&=&  
{\sqrt \frac{m_1 m_2(m_2-m_1 +D_{\nu})} 
{m_3^2 (m_3-D_{\nu})}}e^{-i\phi_1} + 
{\sqrt \frac{m_e(m_{\tau}-D_l)(m_2-m_1 
  + D_{\nu})}{m_{\mu}m_{\tau}m_3}} \nonum \\ 
& &  - 
{\sqrt \frac{m_e(m_{\mu}+D_l)(m_3-D_{\nu})} 
{m_{\mu}m_{\tau}m_3}}e^{i\phi_2}\,, \label{us1} \\
U_{e2} & = & -{\sqrt\frac{m_1}{m_1+m_2 }}e^{-i\phi_1} + 
 {\sqrt \frac{m_e m_2(m_{\tau}-D_l)(m_3-D_{\nu})} 
{m_{\mu}m_{\tau}m_3(m_1+m_2)}} \nonum \\ 
& &  + 
{\sqrt \frac{m_e m_2(m_{\mu}+D_l)(m_2-m_1  
 + D_{\nu})} 
{m_{\mu}m_{\tau}m_3(m_2+m_1)}}e^{i\phi_2}\,, \label{us2} \\
U_{\mu 3} & = &  
-{\sqrt \frac{m_e m_1 m_2(m_2-m_1 +D_{\nu})} 
{m_{\mu}m_3^2 (m_3-D_{\nu})}}e^{-i\phi_1} + 
{\sqrt \frac{(m_{\tau}-D_l)(m_2-m_1  
+ D_{\nu})}{m_{\tau}m_3}} \nonum\\ 
& & - 
{\sqrt \frac{(m_{\mu}+D_l)(m_3-D_{\nu})} 
{m_{\tau}m_3}}e^{i\phi_2}\,.  \label{us3} 
\eeqn } 
The calculations have been carried out for the following 
 cases: $(i)$ both $M_l$ and $M_{\nu}$ real, $M_l$ being diagonal,
$(ii)$ both $M_l$ and $M_{\nu}$ real and non-diagonal,
and $(iii)$ Both $M_l$ and $M_{\nu}$ being complex and non-diagonal. 
To begin with, we  consider the simplest case, wherein  
$M_l$ is diagonal and $M_{\nu}$ is real,   and  
calculate $s_{13}$ corresponding to the different 
SNP solutions and the constraints implied by ANP data. 
For this case,  the mixing matrix $U$  corresponds to 
the  diagonalizing matrix $V_{\nu}$, given in equation 
 (\ref{mnapp}) with $\alpha_{\nu}=\beta_{\nu}=0$.   
Before discussing the results, it is perhaps desirable to discuss some 
of the specific details pertaining to various inputs.
While considering the ANP and SNP constraints, we have taken  
the best fit values corresponding to $\Delta m_{12}^2$ and
$\Delta m_{23}^2$ given in equations (\ref{ans})$-$(\ref{lows}),
however in the case of mixing angles we have considered the ranges
\cite{spec} :
\beqn
     {\rm sin}\,\theta_{23} & = & 0.55 - 0.84,    \label{atdat}  \\
 {\rm sin}\,\theta_{12} & = & 0.42 - 0.71 \qquad  ~~{\rm VO}, 
 \label{vo} \\
 {\rm sin}\,\theta_{12} & = & 0.42 - 0.65 \qquad  ~~{\rm LMA}, 
 \label{lma} \\
 {\rm sin}\,\theta_{12} &=& 0.015 - 0.05 \qquad  ~{\rm SMA}, 
  \label{sma} \\
     {\rm sin}\,\theta_{12} &=& 0.42 - 0.61 \qquad  ~~{\rm LOW},
  \label{low} 
\eeqn
which include the the best fit values given in equations
(\ref{ans})-(\ref{lows}).
 In this case, we have $D_{\nu}$ and one of the neutrino masses as the 
unknown parameters,  the other two masses can be deduced from  
$\Delta m_{23}^2$  and $\Delta m_{12}^2$. 
In principle, the parameter 
 $D_{\nu}$ can take any value, however, in accordance with the similar 
 analysis in the quark sector \cite{p1}, we have restricted its 
 variation within $0 < D_{\nu} < m_3$. 
We choose $m_1$ to be the free parameter and  find 
a range of $s_{13}$ for different values of $m_1$ and $D_{\nu}$, 
 such that 
$s_{12}$ and $s_{23}$ are reproduced within the  SNP and ANP 
constraints. 
In the Table (\ref{tabuniqlma}), we have presented our results 
regarding  $s_{13}$ for different values of 
$m_1$ and $D_{\nu}$ for the LMA solution of SNP. 
For the sake of completeness, 
we have also 
presented in the table the corresponding values of $s_{12}$ and 
$s_{23}$. 
A general survey of the table reveals the following range of $s_{13}$ 
\be s_{13} = 0.04 - 0.13 \,, \label{s13lma} \ee 
 which 
 is in agreement with the range of $s_{13}$ found by Akhmedov 
 {\it et al.} \cite{akho}, for example 
\be s_{13} = 0.05 - 0.15 \,. \ee 
It is interesting to mention the ranges of the neutrino masses 
corresponding to the equation (\ref{s13lma}), for example 
\beqn m_1 &=& (1 - 5) \times 10^{-3}\,{\rm eV}\,, \label{a6} \\ 
 m_2 & = & (6 - 8) \times 10^{-3}\,{\rm eV} \,, \label{b6} \\ 
      m_3 & \simeq &  5 \times 10^{-2}\,{\rm eV} \,, \label{c6} \eeqn 
suggesting  the mass hierarchy  $m_1 \simeq m_2 < m_3$  
for neutrinos. 
In the same vein, it is  interesting to  
 examine the hierarchy pattern of 
the elements of the neutrino mass matrix $M_{\nu}$. In this context, 
it may be noted  that while carrying out the analysis, we 
have varied 
$D_{\nu}$ between  
 0 and  $m_3$, however the allowed range of $D_{\nu}$ is given as 
 $0.24 < \frac{D_{\nu}}{m_3} < 0.6$. This range of 
$D_{\nu}$ helps determining the hierarchy pattern of mass matrix 
elements.   
 As an example,  when the solution 
 corresponding to the row I of the Table (\ref{tabuniqlma})  
 is used to construct 
  $A_{\nu}$,  $B_{\nu}$,  $C_{\nu}$ and  $D_{\nu}$, we obtain 
  the following mass matrix 
 \be 
 M_{\nu}=\left( \ba{ccc}  
 0  & 0.0041 & 0  \\ 
 0.0041 &  0.022     &  0.028  \\  
0 &     0.028    &    0.029    \ea \right) \, . 
 \label{rmm1} 
 \ee 
 The above matrix corresponds to hierarchy  
  $A_{\nu} < B_{\nu} \simeq D_{\nu} \simeq C_{\nu}$, which 
 looks to be somewhat different compared to the hierarchy 
      pattern of quark mass matrix elements.
The mixing matrix corresponding to the above neutrino  
mass matrix  is given as 
\be U = \left(  \ba{ccc}  
  0.903 &     0.426  &    0.050 \\ 
 0.284   &   0.682    &  0.673 \\ 
 0.321  &   0.594  & 0.737 \ea \right).   
\label{rmm2} \ee 
The above mixing matrix is well within the mixing matrix
derived recently by Fukugita and Tanimoto \cite{fuku},
by including the SNO data also.

In a similar manner, one can carry out the  
analysis for the SMA, LOW and VO solutions 
 of SNP. Without presenting the details of results, we summarize the 
 $s_{13}$ ranges evaluated in these cases as 
\beqn s_{13} &=& (0.4 - 2.5)\times 10^{-3}  \qquad ~  {\rm SMA}, \label{x6} \\ 
 s_{13} &=& (1.7 - 6.7)\times 10^{-3}  \qquad ~ {\rm LOW}, \label{y6} \\ 
 s_{13} &=& (0.2 - 12.0)\times 10^{-3}  \qquad  {\rm VO}, \label{z6}\eeqn 
 which very well include the 
 corresponding values given by Akhmedov {\it et al.}, for example 
\be s_{13} \sim  10^{-3}~~({\rm SMA}),~~  
 s_{13} \sim  10^{-2}~~({\rm LOW}),~~ 
 s_{13} \sim  10^{-4} - 10^{-3}~~({\rm VO}). \ee 
The  $m_1,\, m_2$ and $m_3$ ranges corresponding to $s_{13}$, given in 
 equations (\ref{x6})$-$(\ref{z6}),  are 
\beqn m_1 \sim 10^{-8} - 10^{-6}\,{\rm eV},~ &~ 
    m_2 \sim 10^{-3}\,{\rm eV},~&~ 
    m_3 \sim 0.055\,{\rm eV}, \qquad {\rm SMA}, \\ 
  m_1 \sim 10^{-5} - 10^{-4}\,{\rm eV}, ~&~ 
    m_2 \sim 10^{-4}\,{\rm eV},~&~ 
    m_3 \sim 0.055\,{\rm eV}, \qquad {\rm LOW}, \\ 
 m_1 \sim 10^{-5} - 10^{-3}\,{\rm eV}, ~&~ 
    m_2 \sim 10^{-4} \,{\rm eV}, ~&~ 
    m_3 \sim 0.055\,{\rm eV}, \qquad {\rm VO}. \eeqn 
 
 In case of LOW and VO solutions, our conclusions regarding 
 the mass hierarchy as well as  
 the hierarchy of mass matrix elements remain the same as that for the 
 LMA case discussed above, however for the SMA case, we find that the 
 allowed values of $m_1$  are such that the corresponding mass 
 hierarchy is  
$m_1 \ll m_2 < m_3. $

Having discussed the case with diagonal $M_l$, it is interesting to
examine the implications of non-diagonal $M_l$ for $s_{13}$. 
With non-diagonal and real $M_l$ and $M_{\nu}$, 
 the $U_{e2},\, U_{e3}$ and $U_{\mu 3}$ are
 given by equations (\ref{us1})-(\ref{us3}) 
with $\phi_1=\phi_2=0$.  
In this case, apart from $m_1$ and $D_{\nu}$, we have $D_l$ as another 
free parameter. Again, we  find a range of $s_{13}$ by varying 
$D_{\nu}$ within  $0 < D_{\nu} < m_3$  and 
$D_l$ within  $0 < D_l < m_{\tau}$, 
as well as by  
scanning a suitable range of $m_1$ so that $s_{23}$ and 
$s_{12}$ are within the limits given by ANP and SNP solutions.  
In the Table (\ref{tableplma}), we have presented our results 
regarding $s_{12}$,  $s_{23}$ and  $s_{13}$ for different values of 
$m_1$, $D_{\nu}$ and $D_l$ for the LMA solution. 
 The range of $s_{13}$ for the LMA solution is given as  
\be s_{13}~~=~~ \sim 0 - 0.19 \,. \label{lmas13a} \ee 
The corresponding allowed ranges of $m_1$, $m_2$ and $m_3$  are 
\beqn m_1& =& (0.7 - 10.0) \times 10^{-3}\,{\rm eV}, \label{al6} \\ 
 m_2 & = & (6.5 - 11.9) \times 10^{-3} \,{\rm eV}, \label{bl6} \\ 
  m_3 & = &  (5.5 - 5.6) \times 10^{-2}\,{\rm eV}, \label{cl6} \eeqn 
 which are within the ranges found recently
by Osland and Wu \cite{oswu} by considering similar texture for
neutrino mass matrix.

Following a similar procedure as discussed above for LMA case,
 the ranges of $s_{13}$ obtained with the SMA, LOW and VO 
 solutions are  given as 
\beqn s_{13} &=& 0.038 - 0.093 \qquad  {\rm SMA}, \label{xl6}\\ 
 s_{13} &=& 0.042 - 0.095  \qquad  {\rm LOW}, \label{yl6}\\ 
 s_{13} &=&  0.038 - 0.096 \qquad  {\rm VO}. \label{zl6} \eeqn 
The corresponding allowed ranges of $m_1,\, m_2$ and $m_3$ in 
each case are  
\beqn m_1 \sim 10^{-8} - 10^{-5}\,{\rm eV}, ~&~ 
    m_2 \sim 10^{-3}\,{\rm eV},~&~ 
    m_3 \sim 0.055\,{\rm eV}, \qquad {\rm SMA}, \label{smahier} \\ 
  m_1 \sim 10^{-5} - 10^{-4}\,{\rm eV}, ~&~ 
    m_2 \sim 10^{-4}\,{\rm eV},~&~ 
    m_3 \sim 0.055\,{\rm eV}, \qquad {\rm LOW}, \\ 
 m_1 \sim 10^{-5} - 10^{-3}\,{\rm eV}, ~&~ 
    m_2 \sim 10^{-4} \,{\rm eV}, ~&~ 
    m_3 \sim 0.055\,{\rm eV}, \qquad {\rm VO}. \eeqn

The ranges of $s_{13}$ given in  equations (\ref{lmas13a}), 
(\ref{xl6})$-$(\ref{zl6})
 look to be vastly different from the ranges 
 given in equations (\ref{s13lma}), 
 (\ref{x6})$-$(\ref{z6}) evaluated using 
 diagonal $M_l$. This warrants a more detailed comparison of the two 
 cases. 
 One finds that in the case of LMA solution, the values 
of $s_{13}$ obtained are in the same range as obtained 
 with  diagonal $M_l$, however 
the spread is larger. While, 
in the case of solutions pertaining to SMA, LOW and VO solutions 
of SNP,  $s_{13}$  is much higher compared to 
that of the case with diagonal $M_l$. 
 Further, the three 
solutions more or less have same range of values.  
This can be easily understood by 
analyzing the expression for $s_{13}$, which can be written as 
\be s_{13}=s_{13}^d + 
{\sqrt \frac{m_e(m_{\tau}-D_l)(m_2-m_1 
  + D_{\nu})}{m_{\mu}m_{\tau}m_3}}  - 
{\sqrt \frac{m_e(m_{\mu}+D_l)(m_3-D_{\nu})} 
{m_{\mu}m_{\tau}m_3}}\,, \label{s13r6} \ee 
where $s_{13}^d$ is the expression for $s_{13}$ with diagonal $M_l$ 
  and is given as  
\[ s_{13}^d = {\sqrt \frac{m_1 m_2(m_2-m_1 +D_{\nu})} 
{m_3^2 (m_3-D_{\nu})}}\,. \] 
A closer scrutiny of (\ref{s13r6}) reveals that for LMA solution all 
the three terms are of the same order leading to wider spread for 
$s_{13}$. 
For the SMA, LOW and VO solutions, we find that 
the contribution of $s_{13}^d$ is very small compared to the other two 
terms  which are two orders of magnitude 
larger pushing $s_{13}$ up compared to the case with diagonal 
$M_l$.

Important conclusions can be derived from above results regarding 
$s_{13}$.   
For example, in case $s_{13}$ is found to be outside the range 
(0.01-0.1), 
then the solutions SMA, LOW and VO  look to be ruled out for texture 4 
zero mass matrices. Further, it is interesting to emphasize that 
several present analyses, excluding as well as including SNO,
 favor LMA
and LOW solution of SNP, LMA being the preferred solution. The 
present calculations look to be favoring LMA solution as it is able to 
encompass the entire possible range of $s_{13}$.

The hierarchy pattern of the neutrino masses and of the mass matrix 
elements remains more or less  
similar to the  case with diagonal $M_l$. 
For example, for LMA solution 
 the mass matrix  corresponding to the first row of Table 
(\ref{tableplma}) is given as 
 \be 
 M_{\nu}=\left( \ba{ccc}  
 0  & 0.0069 & 0  \\ 
 0.0069 &  0.006     &  0.021  \\  
0 &     0.021    &    0.047    \ea \right) \, . 
 \label{rmma1} 
 \ee 
The  corresponding mixing 
 matrix  is given as 
\be U = \left(  \ba{ccc}  
  0.816 &     0.578  &    0.0002\\ 
 0.473   &   0.668    &  0.574 \\ 
 0.332  &   0.468  & 0.818 \ea \right).   
\label{rmma2} \ee 
The above mixing matrix is again within the mixing matrix
derived by Fukugita and Tanimoto \cite{fuku}.

Recently several authors have discussed the possibility of observing 
CP violation in the leptonic sector
\cite{cplep}-\cite{kangcp},\cite{fuku}.
To include CP violation, 
the corresponding mass matrices $M_l$ and $M_{\nu}$ have to 
be complex. This necessitates  non-zero 
 phases $\alpha_{l(\nu)}$ and $\beta_{l(\nu)}$ in the mass matrices 
given in equation (\ref{nmm6})  
 and the corresponding simplified expressions for
 $U_{e3}$, $U_{e2}$ and $U_{\mu 3}$ are given by the
equations (\ref{us1})-(\ref{us3}).  
Again, following the same procedure as detailed in the previous 
sections, and by scanning the  range of phases $\phi_1$ and 
$\phi_2$ from 0 to $\pi$, 
 we have evaluated the $s_{13}$ ranges corresponding to the 
four SNP solutions and satisfying ANP and CHOOZ 
 constraints, for example 
\beqn s_{13} & = & \sim 0 - 0.20 \qquad\qquad ~{\rm LMA} \\ 
 s_{13}&  = & 0.038 - 0.094 \qquad ~~~{\rm SMA} \\ 
 s_{13}&  = & 0.028 - 0.099\qquad ~~~{\rm LOW} \\ 
 s_{13}&  = & 0.035 - 0.098 \qquad ~~~{\rm VO} \eeqn 
Comparing with the corresponding results obtained in the last section  
without CP violation,  
we find that the range of $s_{13}$ is more or less similar to the case 
with real and non-diagonal
 $M_l$ and $M_{\nu}$.  For a given values of neutrino 
 masses, $D_{\nu}$ and $D_l$, the  $s_{13}$ ranges  
for SMA, LOW and VO solutions show marginal increase 
  with the 
 introduction of phases 
compared to the previous case, however for LMA $s_{13}$ goes up 
 significantly, 
 as is evident 
 from the results presented in the Table (\ref{tabs13}).

As a rough measure of the CP violating phase in the 
 leptonic sector, 
several authors have estimated the maximum value of J$^l_{CP}$ 
\cite{kangcp,jarl}, the 
Jarlskog's rephasing invariant parameter in the leptonic sector. In 
this context, we have also calculated J$^l_{CP}$ from the mass matrices 
using the following expression: 
 \begin{eqnarray} 
       {\rm Det}\,[M_lM_l^{\dagger},M_{\nu}M_{\nu}^{\dagger}] & = & 
      -2i\,{\rm J}^l_{CP}\,(m_{\tau}^2-m_{\mu}^2)(m_{\mu}^2-m_{e}^2) 
       (m_e^2-m_{\tau}^2) \nonumber\\ 
           &  & \mbox{} \times(m_3^2-m_2^2) 
         (m_2^2-m_1^2) 
             (m_1^2-m_3^2)\,. \label{comm6}  \end{eqnarray} 
Varying  phases $\phi_1$ and  $\phi_2$ from 0 to $\pi$, with 
 other 
inputs being the same as used in previous sections, we obtain for LMA, 
SMA, LOW and VO cases  
\beqn {\rm J}^l_{CP} &  < &  0.099 \qquad \qquad~ {\rm LMA}, \\ 
  {\rm J}^l_{CP} &  < &  0.0012 \qquad \qquad{\rm SMA}, \\ 
 {\rm J}^l_{CP} &  < &  0.043 \qquad \qquad ~{\rm LOW}, \\ 
 {\rm J}^l_{CP} &  < &  0.045 \qquad \qquad~{\rm VO}, \eeqn  
which are in agreement with most of the contemporary analyses 
 \cite{jlep,kangcp,fuku}.

To summarize the conclusions, we have considered lepton mass matrices
similar to the texture 4 zero mass matrices, known to be quite
successful in explaining the CKM phenomenology, for finding the 
mixing matrix element
$U_{e3}$ respecting the CHOOZ constraint
with
  $s_{12}$ and $\Delta m_{12}^2$ constrained by SNP and $s_{23}$ and
 $\Delta m_{23}^2$ constrained by ANP. Interestingly,
when the charged lepton mass matrix is taken to be
diagonal,
 the ranges of $s_{13}$ corresponding to different SNP solutions very
 well include the corresponding ranges found by Akhmedov
 {\it et al.} by considering neutrino mass matrices with no texture
zeros. Taking $M_l$ and $M_{\nu}$ both  to be real and
non-diagonal, $s_{13}$ ranges for the four SNP
 solutions come out to be:
$  \sim 0 - 0.19~({\rm LMA}),~ 
 0.038 - 0.093~ ({\rm SMA}),~
  0.042 - 0.095~({\rm LOW}),~
0.038 - 0.096~({\rm VO}).$
Except for LMA case, the other ranges show marked difference compared
to the case with diagonal $M_l$. These conclusions remain
largely valid when the effects of phases are included.
In fact,
 in case $s_{13}$ is found to be outside the range (0.01-0.1), 
then the solutions SMA, LOW and VO  look to be ruled out for texture 4 
zero mass matrices.
  \vskip 0.2cm
{\large\bf Acknowledgements} \\
 The authors would like to thank S.D. Sharma for useful discussions.
 M.R. would like to thank CSIR, Govt. of India for financial support.
 G.A. and M.R. would like to thank the Chairman, Department of Physics
 for providing facilities to work in the department.
M.G. would like to acknowledge the university grant No.
3427/A/8.3.2001.

\begin{table}
\renewcommand{\arraystretch}{1.3} 
\begin{tabular*}{\columnwidth}{@{\extracolsep{\fill}}|ccccc|}  
 \hline\hline 
$m_1 \times 10^{-3}\,{\rm eV}$ &  $\frac{D_{\nu}}{m_3}$ &  
 $s_{12}$ & $s_{23}$ & $s_{13}$  \\ \hline \hline 
 1.3  &    0.4  &   0.43  &   0.67  &   0.05  \\ 
 1.4  &    0.5  &   0.45  &   0.74  &   0.06  \\ 
 1.5  &    0.5  &   0.46  &   0.74  &   0.07  \\ 
 1.6  &    0.5  &   0.47  &   0.74  &   0.07  \\ 
 1.6  &    0.6  &   0.49  &   0.80  &   0.08  \\ 
 1.7  &    0.6  &   0.50  &   0.80  &   0.09  \\ 
 2.1  &    0.6  &   0.54  &   0.80  &   0.10  \\ 
 2.5  &    0.6  &   0.57  &   0.80  &   0.11  \\ 
 3.3  &    0.6  &   0.62  &   0.80  &   0.12  \\ 
 3.4  &    0.6  &   0.63  &   0.80  &   0.12  \\ 
 3.6  &    0.6  &   0.64  &   0.80  &   0.13  \\ 
 3.8  &    0.6  &   0.65  &   0.80  &   0.13  \\ 
 3.9  &    0.5  &   0.63  &   0.74  &   0.11  \\ 
 4.0  &    0.5  &   0.63  &   0.74  &   0.11  \\ 
 4.4  &    0.5  &   0.65  &   0.74  &   0.12  \\ 
\hline \hline 
\end{tabular*} 
\caption{Calculated values of $s_{12}$, $s_{23}$ and $s_{13}$ for the LMA 
solution of SNP for real $M_{\nu}$ and  $M_l$, with $M_l$
 being diagonal.} 
\label{tabuniqlma} 
\end{table}

\begin{table}
\renewcommand{\arraystretch}{1.3} 
\begin{tabular*}{\columnwidth}{@{\extracolsep{\fill}}|cccccc|} 
\hline\hline 
$m_1 \times 10^{-3}\,{\rm eV}$ &  
 $\frac{D_{\nu}}{m_3}$ &  $\frac{D_l}{m_{\tau}}$ &  
 $s_{12}$ & $s_{23}$ & $s_{13}$  \\ \hline \hline 
 5.0   &    0.1  &    0.7  &   0.58  &   0.57   &   0.0002  \\ 
 8.2   &    0.1  &    0.8  &   0.63  &   0.67   &   0.005  \\ 
 8.6   &    0.2  &    0.9  &   0.64  &   0.68   &   0.010  \\ 
 8.8   &    0.1  &    0.7  &   0.64  &   0.59   &   0.020  \\ 
 7.6   &    0.1  &    0.9  &   0.63  &   0.77   &   0.030  \\ 
 8.4   &    0.2  &    0.8  &   0.64  &   0.58   &   0.040  \\ 
 9.1   &    0.3  &    0.9  &   0.65  &   0.60   &   0.050  \\ 
 2.1   &    0.1  &    0.9  &   0.44  &   0.74   &   0.060  \\ 
 1.3   &    0.7  &    0.1  &   0.43  &   0.59   &   0.139  \\ 
 1.7   &    0.7  &    0.1  &   0.49  &   0.59   &   0.153  \\ 
 2.1   &    0.7  &    0.1  &   0.53  &   0.59   &   0.165  \\ 
 2.2   &    0.7  &    0.1  &   0.54  &   0.59   &   0.168  \\ 
 2.3   &    0.7  &    0.1  &   0.55  &   0.59   &   0.171  \\ 
 3.0   &    0.7  &    0.1  &   0.60  &   0.59   &   0.189  \\ 
 3.4   &    0.7  &    0.1  &   0.63  &   0.59   &   0.199  \\ 
\hline  
\hline \end{tabular*}  
\caption{Calculated values of $s_{12}$, $s_{23}$ and $s_{13}$ for the LMA 
solution of SNP for real and  non-diagonal   $M_l$ and $M_{\nu}$.} 
\label{tableplma} 
\end{table}

\begin{table}
\bc 
\renewcommand{\arraystretch}{1.3} 
\bt{|cccc|} \hline\hline 
& & $s_{13}$ & \\ \hline \hline 
& \bt{c}Diagonal $M_l$\\ real $M_{\nu}$ \\ \et & 
 \bt{c} Non-diagonal and\\ real $M_l$ and $M_{\nu}$ \\ \et & 
 \bt{c} Non-diagonal and \\ complex $M_l$ and $M_{\nu}$ \\ \et \\  
 \hline \hline 
LMA & 0.0628 & $ 0.0033$ & $0.0033-0.1422$ \\ 
SMA & $0.0006$ & 0.067 & 0.067 - 0.071 \\ 
LOW & $0.0021$ & 0.067 & 0.067 - 0.083 \\ 
VO  & $0.0002$ & 0.069 & 0.069 - 0.081 \\ 
\hline \hline 
\et 
\caption{The values of $s_{13}$ obtained with fixed 
$m_1, D_{\nu}$ and $D_l$ , with
$\phi_1$ and $\phi_2$ varying from 0 to $\pi$.} 
\label{tabs13} 
\ec  
\end{table} 
 \end{document}